\documentstyle[prl,aps,epsfig,floats,psfrag,axodraw,graphicx]{revtex}
\begin{document}
\draft
\newcommand{\be}{\begin{equation}}\newcommand{\ee}{\end{equation}}
\newcommand{\bea}{\begin{eqnarray}}\newcommand{\eea}{\end{eqnarray}}
\def\no{\nonumber}
\def\eq#1{eq. (\ref{#1})}\def\eqeq#1#2{eqs. (\ref{#1}) and  (\ref{#2})}
\def\lsim{\raise0.3ex\hbox{$\;<$\kern-0.75em\raise-1.1ex\hbox{$\sim\;$}}}
\def\gsim{\raise0.3ex\hbox{$\;>$\kern-0.75em\raise-1.1ex\hbox{$\sim\;$}}}
\def\slash#1{\ooalign{\hfil/\hfil\crcr$#1$}}
\def\order#1{{\mathcal{O}}(#1)}
\def\qbq{q\bar{q}}\def\cbc{c\bar{c}}\def\qbar{\bar{q}}
\twocolumn[\hsize\textwidth\columnwidth\hsize\csname@twocolumnfalse\endcsname
\rightline{\small UCL-IPT-05-07, LPT-ORSAY-05-45}

\title{Suppressed  decay into open charm for the $Y(4260)$ being an hybrid}
\author{E. Kou$^{1,2}$ and O. Pene$^2$}
\address{$^1$~
Institut de Physique Th\'{e}orique, Universit\'{e} Catholique de Louvain, B1348, BELGIUM \\
$^2$~Laboratoire de Physique Th\'{e}orique
Universit\'{e} de Paris-Sud XI 91405 Orsay, FRANCE; UMR8627 CNRS\\}
\date{\today}
\maketitle
\begin{abstract}
We investigate the $Y(4260)$ resonance  recently discovered by the Babar 
collaboration.  We propose the observation of its decay into
$J/\psi \pi \pi$ and its non observation into open charm as a consequence
of it being a charmonium hybrid state with a magnetic constituent gluon. 
We prove a selection rule forbidding its decay into two $S$-wave charmed
mesons in any potential model. We suggest a generalisation of the selection 
rule based only on the heavy quark nature of the charm.    
\end{abstract}
]
\section{INTRODUCTION}
The recent observation~\cite{Aubert:2005rm} of the   $Y(4260)$ resonant
structure  in the $\pi^+\pi^- J/\psi$ recoil mass of the process $e^+e^-\to
\gamma_{\mbox{\tiny ISR}} \pi^+\pi^- J/\psi$ ($\gamma_{\mbox{\tiny ISR}}$:
initial state radiation), is identified as a $J^{PC}=1^{--}$ single
resonance with its mass centred around $\sim 4.26$ GeV and with a width 
$\sim 50$  to 90 MeV. 
While the mass of this resonance is well above the $D\overline{D}$ threshold, 
it has been observed only in the decay process $\pi^+\pi^- J/\psi$.  
It has therefore been claimed that it is not a standard  charmonium state but rather an
exotic.

We will argue in the following that these mysterious
features {\it may be an indication that the $Y(4260)$ is a hybrid
 charmonium in the $1^{--}$ state containing a pseudoscalar colour-octet
 $\bar c c$ and a magnetic constituent gluon:
  a selection rule strongly lessens  its decay to ground 
 state open charm states $D^{(\ast)} \overline D^{(\ast)}$}. In the following 
 we will designate our candidate magnetic hybrid by $H_B$.

A four quark model has been proposed for the $Y(4260)$~\cite{Maiani:2005pe}.
A possible interpretation as a conventional charmonium is studied in~\cite{Llanes-Estrada:2005hz}. 
In~\cite{Zhu:2005hp} Zhu examines several hypotheses and finally favours 
the hybrid interpretation. We  will here
support this opinion by several important dynamical arguments.  
Hybrid states ($\qbq +g$ hadron) are one of the most promising new species of
hadrons.  While the hybrids with the exotic quantum numbers 
such as ($0^{+-}, 1^{-+}, 2^{+-}$) would be observed
as a very striking signal, the other kinds could also be distinguished from the
conventional hadrons by the characteristics of their decay processes.
However care has to be taken about  possible mixing between the latter hybrids  
and conventional states.  Extensive
investigations in searching for the hybrid states have been pursued especially,
in the light hadrons, though no evidence has been confirmed. Now that more and
more new charmed hadrons have been discovered by B factory experiments, 
the hope of discovering  $\cbc +g$ hybrids has raised.  The
spectroscopy of the hybrid states which contain a constituent gluon would 
hopefully unveil some new features of QCD.

We will use the language of the constituent model (a generalisation of  the
quark model with constituent gluons). Hybrid charmonium   is a bound state of 
$c\bar{c}$ and a gluon.  Defining $l_g$ (the relative orbital momentum between
$c\bar{c}$ and $g$), $l_{c\bar{c}}$ (the orbital momentum of $c\bar{c}$ state),
$s_{c\bar{c}}$ (the spin of $c\bar{c}$), the quantum numbers of the hybrid
mesons are:   
\begin{equation} P= (-1)^{l_g+l_{c\bar{c}}}, \ \ \ \ \ \
C=(-1)^{l_{c\bar{c}}+s_{c\bar{c}}+1} \end{equation} 
Thus, a $1^{--}$ state can
be composed either by $(l_g, \ l_{c\bar{c}},\ s_{c\bar{c}})=(0,\ 1,\ 1)$ or  by
$(1, \ 0, \ 0)$. The former possibility has already been studied in~\cite{IPS} and
found that it may not exist as a resonance since it is too strongly 
coupled to the continuum $D\overline{D}$ channels (its estimated width exceeds 
1 GeV). At the same time, it was shown  in
\cite{IPS} within a harmonic oscillator  potential model that the case $(l_g,
\ l_{c\bar{c}},\ s_{c\bar{c}})=(1,\ 0,\ 0)$, which is $H_B$,  obeyed  a strict selection rule
forbidding its decay to  any two $S$-wave final 
mesons. Looking at its wave function one sees that it is proportional to the constituent gluon momentum in cross product with its polarisation (times a scalar function of the momenta). This indicates that we are dealing with a magnetic gluon. 
The same selection rule had been advocated for light 
quarks~\cite{LeYaouanc:1984gh} - \cite{Iddir:1988jd}. This result
has been generalised for light quarks to a more general
potential~\cite{Iddir:2002rf,Iddir:2004ry}. 
This very powerful selection rule results in an important decay pattern of $H_B$ 
which we will discuss in this letter;  the lowest possible open charm
final state comes from  $D^{**}\overline D$, whose threshold is just above the 
observed resonance. 
As a result, we find 
i) Connected diagrams, fig. 1, can be significantly suppressed; 
ii) it is a resonant state with a moderate decay width contrarily to 
the above-mentioned $(l_g, \ l_{c\bar{c}},\ s_{c\bar{c}})=(0,\ 1,\ 1)$.

Concerning the mass of the hybrid states,  a thumb rule tells 
that the constituent gluon is expected to add $0.7 \sim 1$ GeV
to the corresponding quarkonia and the excited hybrid states lie another
 $0.4$ GeV above, which sums up to the statement that the mass of 
 this state would be around $4.2 \sim 4.5$ GeV. 

This thumb rule agrees qualitatively with the outcome of the flux-tube model
for hybrid states~\cite{flux-tube}, as well as the lattice QCD
simulations~\cite{lattice}. The selection rule considered in this paper
has been claimed to be rather general, including flux tube 
models~\cite{Close:1994hc,Page:1996rj}. 
A similar selection rule was also used to account for the missing charm puzzle 
of B decay~\cite{Chiladze:1998ti}.

The proof of the above-mentioned selection rule is of course crucial.
 Within the  simple  chromo-harmonic oscillator model it is shown 
in~\cite{IPS}.  This model is not realistic but it is a convenient toy model. 
We give a short reminder of the model in the appendix, where one can also find
a more concrete description of  our $\cbc g$ hybrid state.

 In the body of this letter we will  consider three issues. In the next section
we will demonstrate  generally the selection rule in a potential model. Next we
will consider the  production mechanism of the $H_B$ and its decay into $J/\psi
\pi\pi$. Finally we will  discuss the corrections to the selection rule. 

\section{Selection rule forbidding $H_B \to D^{(*)} \overline{D}^{(*)}$}
\label{selection}
To lowest order, the decay of the hybrid state is described by the matrix
 element of the QCD interaction Hamiltonian between a hybrid wave function and
 a final state two-meson wave function.  The result in the non-relativistic
 limit is given as factorised  in terms of the colour, spin, spatial and
 flavour overlaps. In the following, we investigate the decay of the hybrid  state  into two  charmed meson states ($H_B \to D^{(*)} \overline{D}^{(*)}$)  through connected diagrams, see fig.~1.   

The simplest interpolating field for the $H_B$ is
\be \bar{c}\gamma_5 \lambda^a B^a_{j}c \ee
 Since the magnetic
field has the quantum number $1^{+-}$, the $c\bar{c}$ forms a pseudoscalar ($0^{-+}$) colour octet.   Thus, the polarisation of the hybrid is found to be
parallel to $B$ field, i.e.,  $\vec{k}\times \vec{A}$. 
The decay into open charm goes through the decay of the spatially polarised 
gluon into an octet spin-one $S$-wave light $q \bar q$ pair and a  recombination  of the two charmed and two light quarks into two mesons, fig.~1.

The colour and isospin overlaps are trivial. The spin overlap leads to a
conservation of the total spin. The hybrid total spin is one (zero
for the $c\bar{c}$ and one for the gluon). The model then forbids the 
decay into $D \overline D$. However the decay into at least one $D^\ast$
($\overline D^\ast$) is allowed by spin conservation. If the final mesons 
are ground state ($D^{(\ast)}, \overline D^{(\ast)} $), parity imposes a 
$P$-wave final state.   
  
Next, we shall describe the spacial part which is at the origin of the 
selection rule we advocate. The spacial overlap is obtained 
as: 
\bea\label{overlap}
&&I=\int \int \frac{d\vec{p_{\cbc}}\ d\vec{k}}{\sqrt{2\omega}(2\pi)6}\ 
\Psi_{l_{H_B}}^{m_{H_B}}
(\vec{p}_{\cbc}, \ \vec{k}) \no \\
&&\Psi_{l_B}^{m_B\ *}(\vec{p}_B)\ 
\Psi_{l_C}^{m_C\ *}(\vec{p}_C) \no \\
&&d\Omega_f Y_l^{m\ *}(\Omega_f)
\eea 
where $\Psi_{l_{H_B}}^{m_{H_B}}$, 
$\Psi_{l_B}^{m_B\ *}(\vec{p}_B)$, and $\Psi_{l_C}^{m_C\ *}(\vec{p}_C)$ 
are the spacial wave functions for the initial hybrid state and the final 
$ D^{(*)}$ and $\overline{D}^{(*)}$ states, respectively. 
The spherical harmonic function $Y_l^{m\ *}(\Omega_f)$ represents the orbital momenta between the two final mesons. 
 We have defined the relative momenta :   
 \bea\label{momenta}
k&=&\frac{(m_c+m_{\bar c})p_g-m_g(p_c+p_{\bar c})}{m_g+(m_c+m_{\bar c})} \no \\
p_{B}&=&\frac{m_{q} p_c-m_c p_{\qbar}}{m_c+m_{q}};\ \ \ 
p_{C}=\frac{m_{q} p_{\bar c}-m_c p_{q}}{m_c+m_{q}}; \nonumber \\ 
p_{\cbc}&=&\frac{p_c-p_{\bar{c}}}{2}; \ \ \ 
p_{\qbq}=\frac{p_q-p_{\bar{q}}}{2}. 
\eea
and in the hybrid state centre of mass system (c.m.s.), we have: 
\be
(p_{\bar{q}}+p_q)=-(p_{\bar{c}}+p_c) = p_g; \ \ \ 
 (p_{\bar{c}}+p_q)=-(p_{\bar{q}}+p_c) \equiv p_f. 
\ee
where $\pm p_f$ are the momenta of the final mesons. 
Note that in the hybrids' c.m.s., we also have: 
\be
p_g=k.
\ee
As a result, we can express all the relevant momenta in terms of $k, 
p_{\qbq}, p_f$: 
\bea\label{pB}
&&p_{\cbc}=p_{\qbq}-p_f, \no \\
&&p_B=-\frac{m_q p_f }{m_q+m_c}+p_{\qbq}-\frac{k}{2}, \ \ \ 
p_C=\frac{m_q p_f }{m_q+m_c}-p_{\qbq}-\frac{k}{2}. 
\eea
Let us  consider the change of variable 
\be\label{symetry}
k \ \to \ -k. 
\ee
keeping $p_{\qbq}, p_{\cbc}, p_f$ unchanged.
We will prove that in the case of $S$-wave final mesons
the overlap integral (\ref{overlap})
changes sign under the change of a variable (\ref{symetry})
and thus must vanish. 
The hybrid wave function is odd in $k$ since $l_g=1$. From formula 
(\ref{pB}) $p_B \leftrightarrow -p_C$. 
In the case of $S$-wave final mesons, the wave functions for $B$ and $C$ in eq. (\ref{overlap}) 
are identical and even in $p_B$ and $p_C$. Their product remains thus unchanged
by the transformation (\ref{symetry}).  The spherical harmonic 
function $Y_l^{m\ *}(\Omega_f)$ is a function of the unit vector $\widehat p_f$
and is thus unchanged. Finally the overlap integral (\ref{overlap}) changes sign
which ends our proof.
 {\it The decay $H_B \to D^{(\ast)}\overline D^{(\ast)}$
 is forbidden in any potential model}.
  
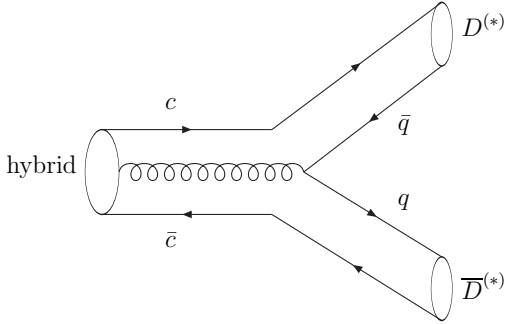
\begin{figure}[h]
\begin{center}
\scalebox{0.8}{
\begin{picture}(200,180)(0,0)
\large 
\Oval(20,80)(20,8)(0)\Gluon(28,80)(115,80){4}{10}
\ArrowLine(20,100)(100,100)\ArrowLine(100,100)(180,160)
\ArrowLine(180,130)(115,80)\Oval(180,145)(15,5)(0)
\ArrowLine(115,80)(180,40)\Oval(180,25)(15,5)(0)
\ArrowLine(180,10)(100,60)\ArrowLine(100,60)(20,60)
\put(50,110){$c$}\put(50,45){$\bar{c}$}
\put(160,100){$\bar{q}$}\put(160,65){$q$}
\put(-25,80){hybrid}\put(190,145){$D^{(*)}$}\put(190,20){$\overline{D}^{(*)}$}
\end{picture}}
\end{center}
\caption{Connected diagram for $H_B \to D^{(\ast)}\overline D^{(\ast)}$. 
This type of process into  $S$-wave final state mesons  
is forbidden for the magnetic constituent gluon by the selection rule. }
\label{fig1}
\end{figure}
\section{Production of the $H_B$ and its allowed decay into $J/\psi \pi\pi$}
The $Y(4260)$ is produced from the
$e^+ e^-$ pair in BABAR experiment~~\cite{Aubert:2005rm}: the virtual photon 
produces a $c \bar c$ in the same quantum numbers as a $J/\psi$: a $1^{--}$
colour singlet. The hybrid 
state is created from two diagrams; the $c\bar{c}$ pair with a gluon emission
from $c$ and $\bar{c}$.  
These two diagrams do not cancel. 
The standard QCD quark-quark-gluon interaction writes
\bea\label{ccg}
\bar c \gamma_i \lambda^a A^i_a c = 
\frac {-i}{2 \, m_c}\; \bar c \,\sigma_{ij}  \lambda^a A^i_a p_g^j c + \cdots
\eea
The emission of a magnetic gluon along the line  
will flip the spin of the charmed quarks from spin 1 (vector) 
 down to spin 0 (pseudoscalar),
and their colour from singlet to octet. {\it The final state has thus exactly the
quantum number of the $H_B$}. This transition is suppressed by one power of the
charm mass. 

{\it A very similar mechanism generates $H_B \to J/\psi \pi\pi$ decay}.
 The emission of an 
additional magnetic gluon from charmed quarks is done via the same (\ref{ccg})
interaction (see fig. 2). The created magnetic gluon can obviously combine with the 
constituent gluon to produce a $0^{++}$ two-gluon state which decays into 
two pions in a  $0^{++}$ state. The charmed quarks have their spin flipped
by the $\sigma_{ij}$ matrix leading to a charmonium state. 
This decay is suppressed by one power of the charm mass but has a large
available phase space which may explain the significant branching ratio
observed in experiment.

\begin{figure}[h]
\begin{center}
\scalebox{0.8}{
\begin{picture}(200,180)(0,0)
\large 
\Oval(20,110)(20,8)(0)\Gluon(28,110)(92,96){4}{7}
\ArrowLine(20,130)(100,130)\ArrowLine(100,130)(180,160)
\Oval(180,140)(20,8)(0)
\ArrowLine(180,120)(120,90)\ArrowLine(120,90)(20,90)
\Gluon(88,96)(130,40){4}{8}
\Gluon(100,130)(135,56){4}{10}
\ArrowLine(160,70)(180,70)\ArrowArcn(160,40)(30,270,90)\ArrowLine(180,10)(160,10)
\ArrowLine(180,55)(160,55)\ArrowArc(160,40)(15,90,270)\ArrowLine(160,25)(180,25)
\put(50,140){$c$}\put(50,75){$\bar{c}$}
\put(-25,110){hybrid}\put(190,145){$J/\psi$}\put(190,60){$\pi$}\put(190,15){$\pi$}
\end{picture}}
\end{center}
\caption{Decay $H_B \to J/\psi \pi \pi$ for hybrid with 
the magnetic constituent gluon. } 
\label{fig2}
\end{figure}
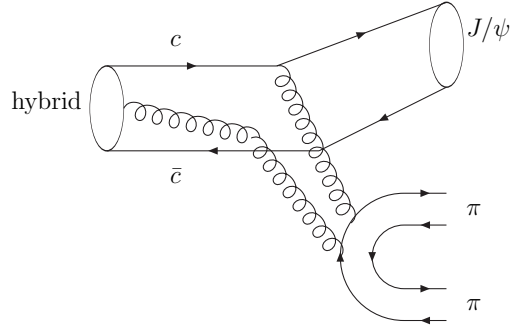
\section{Corrections to the selection rule}
One correction comes via a mixing with a standard charmonium state. 
Indeed the interaction (\ref{ccg}) also induces a mixing of the hybrid with
neighbouring excited charmonia such as $\psi(4160)$. However this interaction is
${\cal O}(1/m_c)$
suppressed and furthermore these excited states have many 
nodes on their wave functions so that the overlap with the $c \bar c$ $0^{-+}$
octet is expected to be rather small. Since the charmonium excitations are
allowed to decay into $D^{(\ast)}\overline D^{(\ast)}$, this mixing induces 
a small correction to the selection rule: it is a second order mechanism 
which implies an additional factor $p_g$, see (\ref{ccg}), which invalidates the
argument of section  \ref{selection}.

Relativistic effects on the charmed quarks may induce other  ${\cal O}(1/m_c)$
corrections to the selection rule since it has only be proven in the
non-relativistic framework.

There are also relativistic corrections 
related to the light quarks, which are difficult to estimate.
However, there is a general argument which implies that they should 
also be suppressed in the infinite $m_c$ limit. Following the philosophy
of the HQET, we consider separately the dynamics of the heavy quarks and
 that of the light quanta (gluons, light quarks). The initial state has        
$c \bar c$ in an $S$-wave, and the light quanta (constituent gluon) possess
an orbital excitation relative to the heavy quark system. 
The final state contains an orbital momentum between the two heavy quarks 
if we consider $S$-wave final mesons. The orbital excitation has to be
transferred from the light system, the ``brown mock'', to the heavy quark 
system. This is presumably suppressed for the following reason. 
The heavy quark system has a vanishing spatial size in the infinite mass
 limit. The cloud of light quanta has a constant size. The overlap vanishes
 in this limit. We can then argue that the orbital momentum transfer is
 consequently suppressed. Of course this is a qualitative argument which 
 needs to be demonstrated and checked. But we believe it to be reasonably 
 convincing.  
 
\section{Other decay channels}

Beyond the violation of the selection rule which should allow a non negligible
branching fraction for the  $Y(4260)\to D^{(*)}\;\overline{D}^{(*)}$ channel,
we must also consider  $Y(4260)\to D^{**}\;\overline{D}^{(*)} \to
D^{(*)}\overline{D}^{(*)} \pi's$ and $Y(4260)\to D^{(*)}\;\overline{D}^{**} \to
D^{(*)}\overline{D}^{(*)} \pi's$. They are not forbidden by the selection rule.
The $Y(4260)$ lies  below the $D^{**}\;\overline{D}^{(*)}$ thresholds (although
very close to the $D_1(2420) \overline D$ one). But these resonances are not
narrow and the decay via a virtual $D^{**}\;\overline{D}^{(*)}$ should not be
small. Therefore we would expect the  dominant decay channel to be $Y(4260)\to
D^{(*)} \overline{D}^{(*)}\pi's$ which might explain a width as large as 90 MeV
for the $Y(4260)$.

\section{Conclusions}
We have argued that the recently observed $Y(4260)$ shows peculiar 
characteristics possessed by a new type of hadrons, an hybrid state,  namely, a bound state of an octet $0^{-+}$ $\cbc$ state and gluon in a $P$-wave (a magnetic gluon), that we call $H_B$.  The decay of $H_B$ is restricted by an important selection rule: the symmetries of the wave functions 
forbid the decay  into two $S$-wave open charm  final mesons in any potential model.
Therefore, $H_B$ cannot  decay into  e.g. $D^{(*)}\overline{D}^{(*)}$ and thus,
has a relatively narrow width, which matches the experimental observation for 
$Y(4260)$. We also showed that the observation channel $Y(4260)\to J/\psi \pi
\pi$ can be naturally explained by $H_B$. At the same time we expect decays such 
as $Y(4260)\to D^{**}\;\overline{D}^{(*)} \to D^{(*)}\overline{D}^{(*)} \pi's$
via a virtual $D^{**}$  to be dominant and
 $Y(4260)\to D^{(*)}\;\overline{D}^{(*)}$ not to be negligible via a violation 
 of the selection rule.
The $H_B$ mass is roughly estimated
to be around $4.2 \sim 4.5$ GeV, though its rigourous prediction would be an
interesting challenge for the lattice QCD. We have discussed the mixing with 
ordinary charmonia and the relativistic corrections. We argue that the latter
are ${\cal O}(1/m_c)$ suppressed using a HQET inspired argument.  A deeper 
theoretical understanding of these issues is needed as well as a search for 
further predictions concerning the properties of $H_B$ to be confronted 
with experimental data concerning the $Y(4260)$ or other similar resonances, 
such as $X/Y(3940)$~\cite{Abe:2004zs}. In particular, several other hybrids
protected by the same selection rule are expected which should be compared
with the increasing number of resonant candidates in this region. Of course, 
all of them are flavor-$SU(3)$ singlets which discriminates clearly  
 this hypothesis from the four quark model. 


\bigskip
\noindent 
{\bf Acknowledgements}\\
The authors would like to acknowledge Guy Wormser for an interesting talk which
 triggered this work. We also thank Ikaros Bigi 
 for interesting discussions  on this issue and Luciano Maiani for stimulating 
questions and suggestions. The work by E.K. was supported by the Belgian
Federal Office for Scientific, Technical and Cultural Affairs through the
Interuniversity Attraction Pole P5/27. 
 
\appendix
\section{}
The $\qbq g$ system can be modelled  by a double chromo-harmonic 
 oscillator~\cite{Iddir:2000yb}:   \be
H_{hyb}= \frac{P^2}{2M}+\frac{p^2_{\qbq}}{2\mu_{\qbq}}+\frac{k^2}{2\mu_g}
-\frac{7b_0}{12}r^2_{\qbq}-3b_0r^2_{H_B} \label{eq:hamiltonian} \ee where $b_0$ is
the level spacing.  The configuration is given as follows; in the centre of
mass system:   \be P=p_q+p_{\qbar}+p_g, \ \ \ M=m_q+m_{\qbar}+m_g \ee the
relative momenta have been defined in eq. (\ref{momenta}). The corresponding
conjugate variables are:
\bea 
&&r_{\qbq}=r_q-r_{\qbar}, \ \ \  \mu_{\qbq}=\frac{m_qm_{\qbar}}{m_q+m_{\qbar}} \no
\\  &&r_{H_B}=r_g-\frac{r_q+r_{\qbar}}{2}, \ \ \ 
\mu_{H_B}=\frac{m_g(m_q+m_{\qbar})}{m_g+(m_q+m_{\qbar})}.
\eea 
  The solution to
the \eq{eq:hamiltonian} can be summarised as:  \be
\Psi^{m_i}_{l_i}(p_i)=\sqrt{\frac{16 \pi^3 R^{2l_i+3}_i}{\Gamma
(\frac{3}{2}+l_i)}}p_i^{l_i}Y^{m_i}_{l_i}(\theta,
\Omega)e^{-\frac{1}{2}R^2_{i}p^2_i} \label{eq:wf} \ee where $i=\qbq, g$ and
$l_{\qbq}$ and $l_g$ are the relative orbital momentum between $\qbq$ and
between $g$ and $\qbq$ centre of mass, respectively.  The radial part of the
solutions are obtained as:  \bea R^2_{\qbq}&=& 1/
\sqrt{2\mu_{\qbq}\left(\frac{-7b_0}{12}\right)} \label{eq:Rq}\\ R^2_{g}&=& 1/
\sqrt{2\mu_g (-3b_0)}\label{eq:Rg} \eea From this result, we can also obtain
the level spacing between the ground state and the first excited state by using
a relation:  \be \omega_i=1/(\mu_i R_i^2).  \ee When considering the heavy
quarks as constituents,    a simple physical picture of such a hybrid state can
be drawn from these expressions. The ratio of  \eqeq{eq:Rq}{eq:Rg}, which
represents the size of the hybrid state  relative to  $\qbq$ state,  \be
\frac{R_g^2}{R_{\qbq}^2}=\sqrt{\frac{7\mu_{\qbq}}{36\mu_{H_B}}} \ee  goes to infinity when  $m_q \to \infty$, so that the heavy quark system shrinks.  In
the same manner, we find  \be \frac{\omega_g}{\omega_{\qbq}}=\sqrt{\frac{36
\mu_{\qbq}}{7\mu_{H_B}}} \ee  expressing that 
the heavy quarks oscillate slowly as compared to the gluon frequency
in this limit.  On the other hand, the charm quark being not so heavy, these ratios are far from large:  \be \frac{R_g^2}{R_{\qbq}^2}\simeq
0.51, \ \ \  \frac{\omega_g}{\omega_{\qbq}}\simeq 2.6 \label{eq:ratio} \ee for
$m_c=m_{\bar{c}}=1.7$ GeV, $m_g=0.8$ GeV.
Therefore, while 
the faster oscillation of $g$ is somehow observed, $c\bar{c}$ is not
 really shrunk.   Indeed, these values are obtained from our potential in
\eq{eq:hamiltonian} containing a certain colour configuration of the states: 
\eq{eq:ratio} is the direct consequence of the fact that the $c\bar{c}$ 
forms an octet state which would fall apart if there was not a screening by the gluon
cloud: the resulting string tension is small. On the contrary the string tension between $\cbc$ and $g$ is large  since it is the string tension between 
two color octets forming a singlet. 
The picture is that of high frequency light quanta and low frequency heavy quarks.  



\end{document}